\documentclass[a4paper,11pt]{article}

\usepackage{pos}
\usepackage{lipsum} %package to generate placeholder text in the following
\usepackage{hyperref}

\usepackage{wrapfig}
\usepackage{multicol}

\usepackage{lineno}
\title{Extending the IceCube search for neutrino point sources in the Northern sky with additional years of data}

\ShortTitle{IceCube Point Source Searches in the Northern Sky}

\author{The IceCube Collaboration \\{\normalsize \normalfont(a complete list of authors can be found at the end of the proceedings)}\\}

\emailAdd{chiara.bellenghi@tum.de}

\FullConference{The 38th International Cosmic Ray Conference (ICRC2023)\\ 26 July -- 3 August, 2023\\ Nagoya, Japan}

\abstract{

The IceCube Neutrino Observatory is a one-cubic-kilometer-sized neutrino telescope deployed deep in the Antarctic ice at the South Pole. One of IceCube’s major goals is finding the origins of astrophysical high-energy neutrinos. In 2022, IceCube identified the strongest point-like neutrino source so far, the active galaxy NGC~1068. Analyzing 9 years of muon-neutrino data from the Northern Sky recorded between 2011 and 2020, the emission from NGC~1068 is significant at 4.2~$\sigma$. We present a planned extension to this search with additional years of data. One of these years includes data from 2010 when IceCube was only partially constructed. We discuss the improvement in sensitivity and discovery potential for neutrino point sources across the Northern sky. We show that by building on the established analysis techniques, previous observations could be improved, not only for NGC~1068 but for all possible sources in the Northern sky.
\vspace{4mm}
{\bfseries Corresponding authors:}
Chiara Bellenghi$^{1*}$, Martin Ha Minh$^{1}$, Tomas Kontrimas$^{1}$, Elena Manao$^{1}$, Rasmus {\O}rs{\o}e$^{1}$, Martin Wolf$^{1}$\\
{$^{1}$ \itshape Technical University of Munich, TUM School of Natural Sciences, Department of Physics, James-Franck-Straße 1, D-85748 Garching bei M\"unchen, Germany}\\[4mm]
$^*$ Presenter

\ConferenceLogo{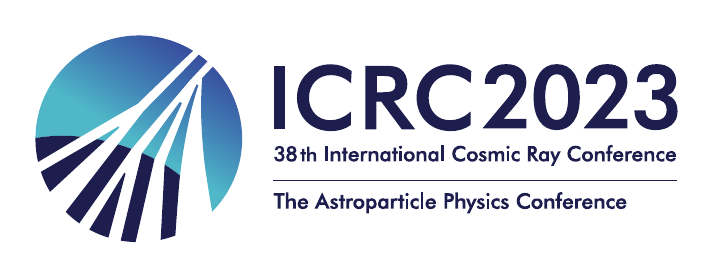}

\FullConference{The 38th International Cosmic Ray Conference (ICRC2023)\\ 26 July -- 3 August, 2023\\ Nagoya, Japan}
}

\begin{document}

\maketitle

\section{Introduction}\label{intro}

\noindent In 2013, IceCube announced the discovery of an astrophysical diffuse flux of high-energy neutrinos \cite{IceCubeScience2013}. Since then, these particles have become a promising probe to solve the long-standing puzzle regarding the origin of high-energy cosmic radiation.
In 2018, the detections of the blazar TXS~0506+056 \cite{TXSScience2018,IceCubeScience2018} marked the identification of the first transient source of high-energy neutrinos, but IceCube has also always been searching for continuous neutrino emission
%from all directions in the sky
(see, \textit{e.g.}, \cite{IceCubeApJ2009,IceCubePRL2020}).
These searches use the information on spatial clustering and energy distribution of the events, both encompassed in an unbinned likelihood-ratio test \cite{BRAUN2008299}, to distinguish between the astrophysical signal from a point-like source and the atmospheric and cosmic background.
The most recent time-integrated search in the Northern sky using ${\sim}9$ years of muon-neutrino track-like events found evidence for neutrino emission from the Seyfert II galaxy NGC~1068 at 4.2$\,\sigma$ \cite{IceCubeScience2022}. This result is the combined product of multiple improvements, including data processing using the latest detector calibration and consistent event filtering, an improved description of the probability density functions (pdfs) in the likelihood using precise numerical estimations obtained from simulations, and new reconstruction algorithms \cite{IceCubeScience2022}. Since the last search, IceCube collected 3 more years of data that can now be used to update this analysis. Additionally, data recorded in 2010 using a partial detector configuration with 79 strings of optical modules are included, extending the livetime by ${\sim}4$ years in total. Besides the increased statistics in the experimental data, we more than doubled the number of simulated events for a more precise modeling of the pdfs at the edges of the parameter space. Through this work, the sensitivity to astrophysical signals will improve, and the statistical significance of the neutrino emission from NGC~1068 is expected to rise. Finally, we investigate the potential of a new energy reconstruction method using Graph Neural Networks.

\section{Event selection and data sample}\label{sec:data_sample}
\noindent The IceCube neutrino telescope, deployed in the deep Antarctic glacier at the South Pole, currently consists of 86 strings, each instrumented with 60 Digital Optical Modules (DOMs) \cite{IceCubeJInst2017}. In 2010, the detector was not yet completed, and IceCube recorded data with a partial configuration, comprising only 79 strings. We will refer to those different detector configurations as IC86 and IC79, respectively.
We plan to extend the previous search for neutrino point sources in the Northern sky \cite{IceCubeScience2022} with several years of data, for an additional livetime of ${\sim}4$ years, of which ${\sim}3.2$ years are from the most recent IC86 data, collected between May 2020 and July 2023, and ${\sim}0.8$ years are from IC79 experimental data. 
The extended data sample totals a livetime of ${\sim}13$ years and contains approximately one million events, corresponding to a ${\sim} 40\%$ increase in statistics.

\noindent The quality of the directional reconstruction of the events is of crucial importance in the search for point-like sources. For this reason, the data sample is constructed by selecting track-like muon events characterized by a linear signature in the detector, thus providing the best pointing capability.
%Another advantage of track-like events is that, being produced by muons that propagate in a straight line for several hundred meters in the ice, they are more likely to lose their energy in the detector volume compared to other event topologies, providing a maximal effective area.
%Another advantage of track-like events is that the neutrino interaction can also happen outside the detector volume and still be detected since the resulting muon travels for several hundred meters in the ice. Hence, the volume where we can detect neutrino interactions is larger than the instrumented volume, providing a maximal effective area.
Since TeV muons can travel several kilometers in the ice, neutrino interactions outside the instrumented volume can still be detected, resulting in a larger effective detection volume.
To increase the purity of the data sample, we limit the search for point-like sources to the Northern sky, where the atmospheric muon background is suppressed by absorption in Earth. The sample used for this analysis contains events with energies above 100~GeV recorded in a declination ($\delta$) range between $-5^\circ$ and $90^\circ$. Here, most of the track-like events are produced by muons induced by the interaction of neutrinos in the Earth. At energies between 100~GeV and 100~TeV, the sample is dominated by events produced by atmospheric neutrinos. At higher energies, the astrophysical neutrino flux starts to emerge. According to the most recent measurement of the astrophysical diffuse neutrino flux \cite{IceCubeDiffuseAPJ2022}, we expect our sample to be composed of ${\sim}0.4\%$ of astrophysical events. 

\noindent For this analysis, the agreement between the simulations and the experimental data in all the interesting observables is vital. The distributions of the simulated events are used to estimate the pdfs for the observables, which are then used for the statistical test (see \autoref{sec:KDE}). For the IC86 detector configuration, the simulations already proved to reproduce the distributions of the experimental data \cite{IceCubeScience2022}. This was not the case for the IC79 detector configuration.

\subsection{The partial IceCube detector configuration: IC79}\label{sec:IC79}

\begin{figure}[t]
    \begin{center}
        \includegraphics[width=\textwidth]{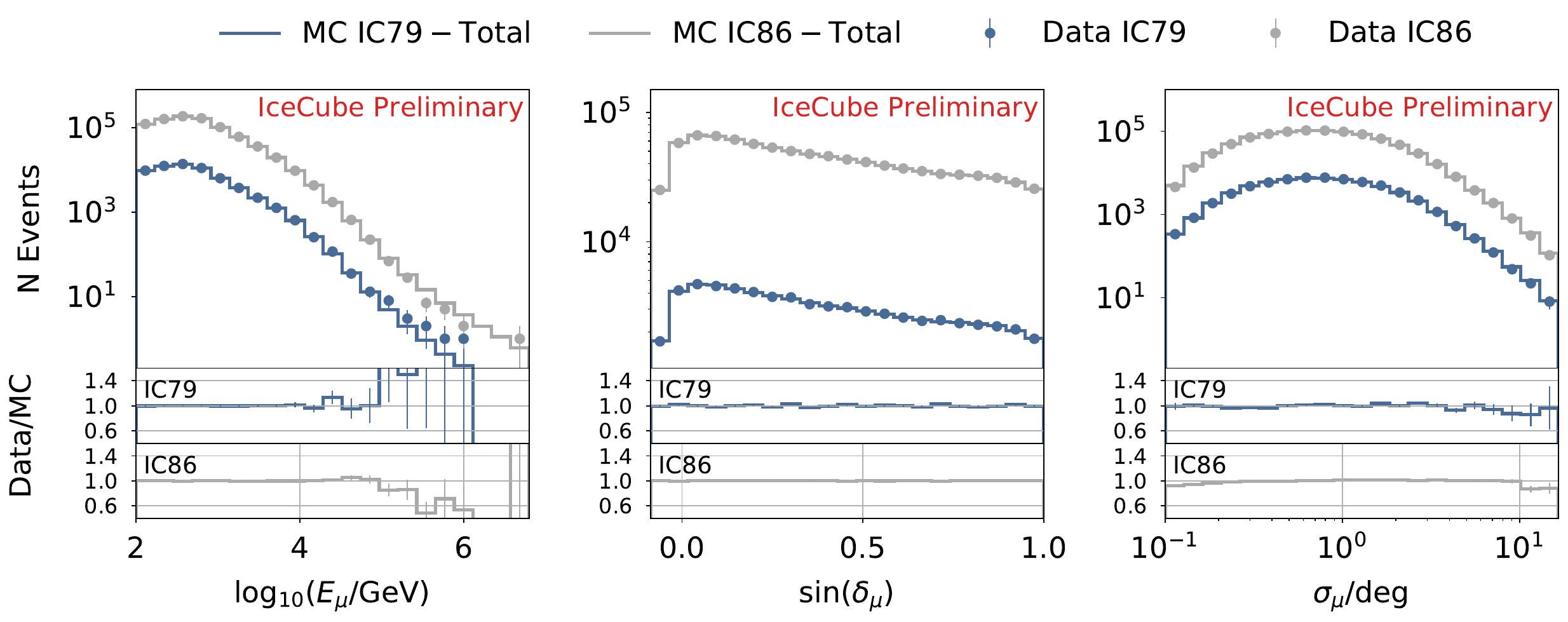}
    \end{center}
    \caption{Agreement between the experimental data (dots) and the simulations (solid lines) in the observables for both IC79 (blue) and IC86 (grey). From left to right, we show the reconstructed muon energy $E_{\mu}$, the sine of the declination $\delta_{\mu}$, and the reconstruction quality estimator on the track direction $\sigma_{\mu}$. In the lower panels, we display the ratios between the experimental data and the simulations for both detector configurations.}
    \label{fig:ic79}
\end{figure}

\noindent Of the seven added strings between IC79 and IC86, two belong to the DeepCore section of the detector \cite{IceCubeJInst2017}, which is optimized for low-energy physics searches and is therefore omitted.
The remaining 5 strings form a line along one of the outer edges of the detector.
%In the previous iteration of this analysis, we found a mismatch between the experimental data and the simulations for the IC86 string configuration, specifically for the reconstructed muon energy and the directional reconstruction quality estimator, as can be seen through the grey ratio lines in \autoref{fig:ic79}.
In the previous iteration of this analysis, IC79 data could not be included for a lack of suited Monte Carlo (MC) simulations. Indeed, the analysis requires $\mathcal{O}$(10M) simulated events to be in good agreement with the observed data to ensure that the observable space is well represented. At the time, such a large set of simulations for the IC79 detector was unavailable, and using the IC86 simulations to describe IC79 resulted in mismatches in the distribution of the observed muon energies. Even though the two detector configurations are minimally different in their effective areas, 
%Upon further investigations on the observed disagreement, we realized that 
the reconstructed muon energy changes for ${\sim}10\%$ of the events depending on which strings detect them.
%, even though the two detector configurations are minimally different in their effective areas.
We use a deep neural network (DNN) \cite{IceCubeScience2022} to reconstruct the energy of the muon track at its first intersection with the detector as a proxy for the energy of the original neutrino. For the same incoming muon, this quantity can be a lower estimate in the IC79 configuration compared to the IC86 one when the missing strings are involved in the detection.
The solution is to procure simulations that accurately represent the IC79 detector configuration.
An efficient way of doing this without generating a large number of simulations from scratch is to take the IC86 simulation sets at the earliest possible processing level, remove the data recorded in the newly added strings, and reapply the full data processing to the final level.
In \autoref{fig:ic79}, we show a comparison between the distributions of the reconstructed muon observables --- energy ($E_{\mu}$), direction given by right-ascension and declination ($\boldsymbol{d}_{\mu} = (\alpha_{\mu}, \delta_{\mu})$), and directional reconstruction quality estimator ($\sigma_{\mu}$) --- of the experimental data and the MC using both, the IC86 and IC79 string configurations. The new IC79 simulations describe the experimental data accurately with maximal discrepancies of a few percent in the high-statistics regions. 
%For the reconstructed muon energy, specifically, the shift to lower energies seen in the experimental data before is now reproduced by the simulation.
As a result, we obtained the agreement between experimental data and MC needed to extract the true distributions of the observables.

\section{Improved modeling of the point spread function}\label{sec:KDE}
\noindent In the search for point-like neutrino sources, we use the unbinned maximum likelihood method described in \cite{BRAUN2008299}. The test statistic (TS) is defined as the negative logarithm of the likelihood-ratio given the background hypothesis of only atmospheric and cosmic neutrinos and the signal hypothesis of astrophysical neutrinos from a point-like source that cluster on top of the background: %($\mu_{\mathrm{ns}} > 0$):
\begin{equation}
    \label{eq:ts}
    \mathrm{TS} \equiv -2 \log\left(\frac{\mathcal{L}(\mu_{\mathrm{ns}}=0\,|\, \boldsymbol{x})}{\sup_{\mu_{\mathrm{ns}},\,\gamma}\mathcal{L}(\mu_{\mathrm{ns}},\,\gamma,\,\boldsymbol{d}_{\mathrm{src}}\,|\,\boldsymbol{x})}\right),
\end{equation}
where $\mu_{\mathrm{ns}}$ is the mean number of signal events, $\boldsymbol{x}$ is the vector of observables $(E_{\mu}, \delta_{\mu}, \sigma_{\mu})$, $\boldsymbol{d}_{\mathrm{src}}$ is the source direction, and $\gamma$ is the spectral index of an assumed power-law spectrum of $\Phi=\Phi_0 \cdot (E/E_0)^{-\gamma}$, with neutrino energy $E$ and flux normalization $\Phi_0$ at some pivot energy $E_0$.
The likelihood function 
\begin{equation}
    \label{eq:final_ps_llh}
    \mathcal{L}\left(\mu_{\mathrm{ns}},\,\gamma \mid \boldsymbol{x} \right) =\prod_{i=1}^N\left\{\frac{\mu_{\mathrm{ns}}}{N} \cdot  f_{\mathrm{S}}\left(\boldsymbol{x}_i \mid \gamma\right)+\left(1-\frac{\mu_{\mathrm{ns}}}{N}\right) \cdot f_{\mathrm{B}}\left(\boldsymbol{x}_i\right)\right\}
\end{equation}
for $N$ events is the weighted sum of a signal  $f_{\mathrm{S}}$ and a background $f_{\mathrm{B}}$ pdf. The signal pdf can be factorized into a spatial pdf term, which describes the spatial clustering of the events around the source, and an energy pdf term, which describes their energy distribution.

% \noindent We use MC simulations with the \textsc{Meerkat} \cite{Poluektov:2014rxa} software package for the numerical kernel density estimation (KDE) of each pdf term comprising the likelihood function because they closely describe the experimental data distributions in the chosen observables for both IC79 (see \autoref{sec:IC79}) and IC86 \cite{IceCubeScience2022} detector configurations.
\noindent Since the simulations closely describe the experimental data distributions of the observables for both detector configurations (see \autoref{fig:ic79}), we can use them to estimate each pdf term in the likelihood function numerically. This is done by using the kernel density estimation (KDE) method with the \textsc{Meerkat} \cite{Poluektov:2014rxa} software package.
Here, we are mostly interested in the spatial pdf term $f_{\mathrm{spat}}\left(\psi\,|\,E_{\mu},\,\sigma_{\mu},\,\gamma \right)$. It describes the point spread function,
% \begin{equation}
%     f_{\mathrm{S}}\left(\psi\,|\,E_{\mu},\,\sigma_{\mu},\,\gamma \right) =
%     \frac{
%         f_{\mathrm{S}}\left( \psi,\,E_{\mu},\,\sigma_{\mu}\,|\,\gamma \right)
%         }{
%         f_{\mathrm{S}}\left( E_{\mu},\,\sigma_{\mu}\,|\,\gamma \right)
%         } ,
%     \label{eq:sig_spatial_pdf}
% \end{equation}
where $\psi$ is the angular distance between the source and the reconstructed muon direction.
Before this approach was conceived, the spatial pdf term was described by a bivariate Gaussian approximation
% \begin{equation}
%     \mathrm{S} \left( E_\mu,\, \psi,\, \sigma\, |\, \theta_s \right) = \frac{1}{2\pi\, \sigma^2} \exp^{-\frac{\psi^2}{2\,\sigma^2}}
% \end{equation}
\cite{BRAUN2008299}. In comparison, the numerically estimated $f_{\mathrm{spat}}$  includes the dependency on the spectral shape of the emission via the spectral index parameter $\gamma$ and describes the distributions of simulated events more accurately, especially at lower energies and for emissions with softer spectral indices.

\begin{figure}[t]
    \centering
    \includegraphics[width=1\textwidth]{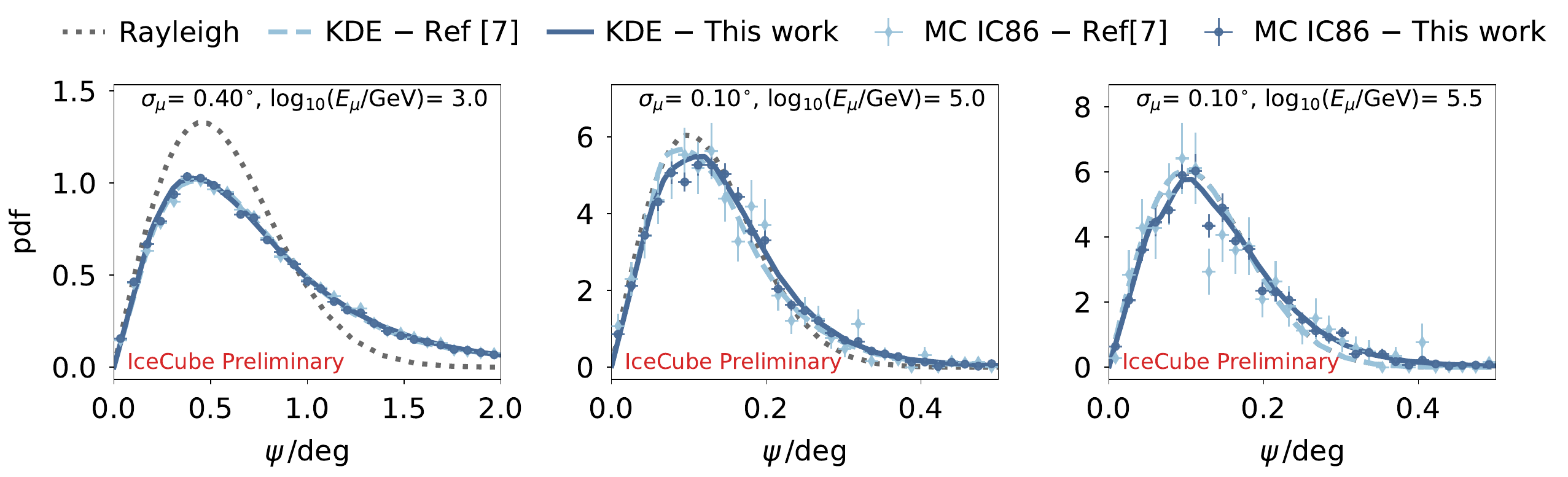}
    \caption{Point spread function examples for $\gamma=3.0$ at different values of the directional reconstruction quality estimator $\sigma_\mu$ and reconstructed muon energy $E_\mu$. In each panel, we show the Rayleigh pdf (projection on the angular distance of the 2D Gaussian), numerically estimated KDE pdfs, and binned MC events. In the rightmost panel, the spatial pdf from Ref.~\cite{IceCubeScience2022} (dashed light blue) falls back to the Rayleigh approximation.}
    \label{fig:spatial_kde}
\end{figure}

\noindent The MC used to construct the KDE pdfs in previous analyses \cite{IceCubeScience2022}
%, later referred to as "MC previous work",
contained only simulations of muon neutrino interactions. In this analysis, we estimate the pdfs from an updated MC sample, containing $\sim$2.5 times more events than the previous work and including muons produced in tau neutrino interactions. The increased statistics of the new MC allow a better description of the point spread function. Especially, we improve the observables parameter space at the edges, where the previous MC did not contain enough simulated events to describe the underlying distributions correctly, and the numerical KDE pdf estimation had to fall back to the analytical approximation. The improvement is shown in \autoref{fig:spatial_kde}.
Even though the experimental data contain only $\mathcal{O}$(100) events at the edges of the parameter space, it is important to model their pdfs precisely. For example, high energy events with good directional reconstruction are less compatible with the background expectation and would strongly influence the signal likelihood term.

\section{Projected analysis performance}\label{sec:performance}

\begin{figure}
    \centering
    \includegraphics[width=0.8\textwidth]{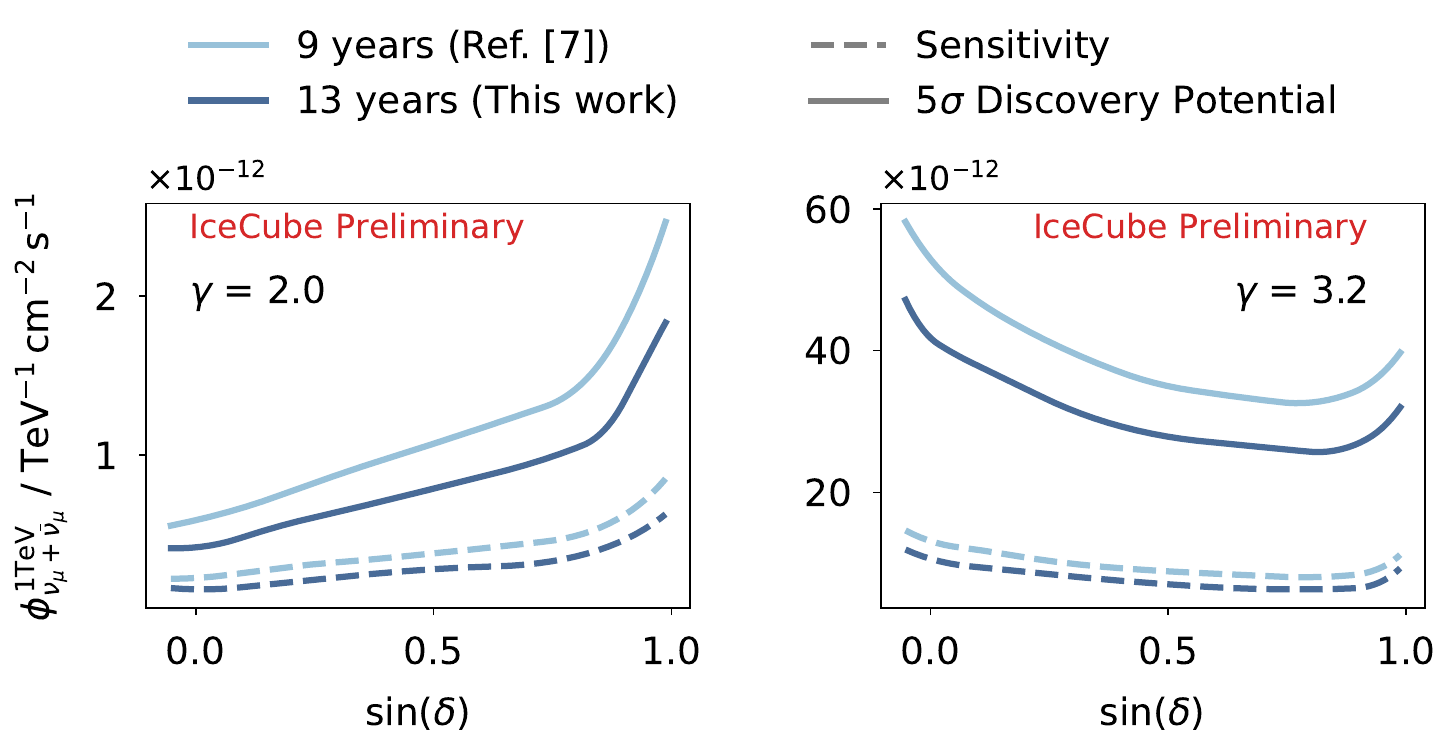}
    \caption{Sensitivity and discovery potential fluxes as a function of the source declination. The left panel shows a hard spectrum of $\gamma=2.0$ and the right one a soft spectrum of $\gamma=3.2$.}
    \label{fig:sensitivity}
\end{figure}

\noindent To estimate the chances of the analysis of detecting an astrophysical flux of high-energy neutrinos and compare to previous studies \cite{IceCubeScience2022}, we use two performance parameters: the sensitivity flux, defined as the flux that would be detected with a TS (see \autoref{eq:ts}) value larger than the median of the TS distribution obtained under the background-only hypothesis with 90\% probability; the 5$\,\sigma$ discovery potential flux, defined as the flux that would be detected with a local significance of 5$\,\sigma$ with 50\% probability.
We calculate these fluxes by generating pseudo-experiments with point sources simulated at various declinations and with spectral indices 2.0 and 3.2. We find that both the sensitivity and the discovery potential improve by ${\sim}30\%$ in the case of hard spectral emission and by ${\sim}20\%$ for steeper energy fluxes (see \autoref{fig:sensitivity}).

\begin{figure}
    \centering
    \includegraphics[width=0.85\textwidth]{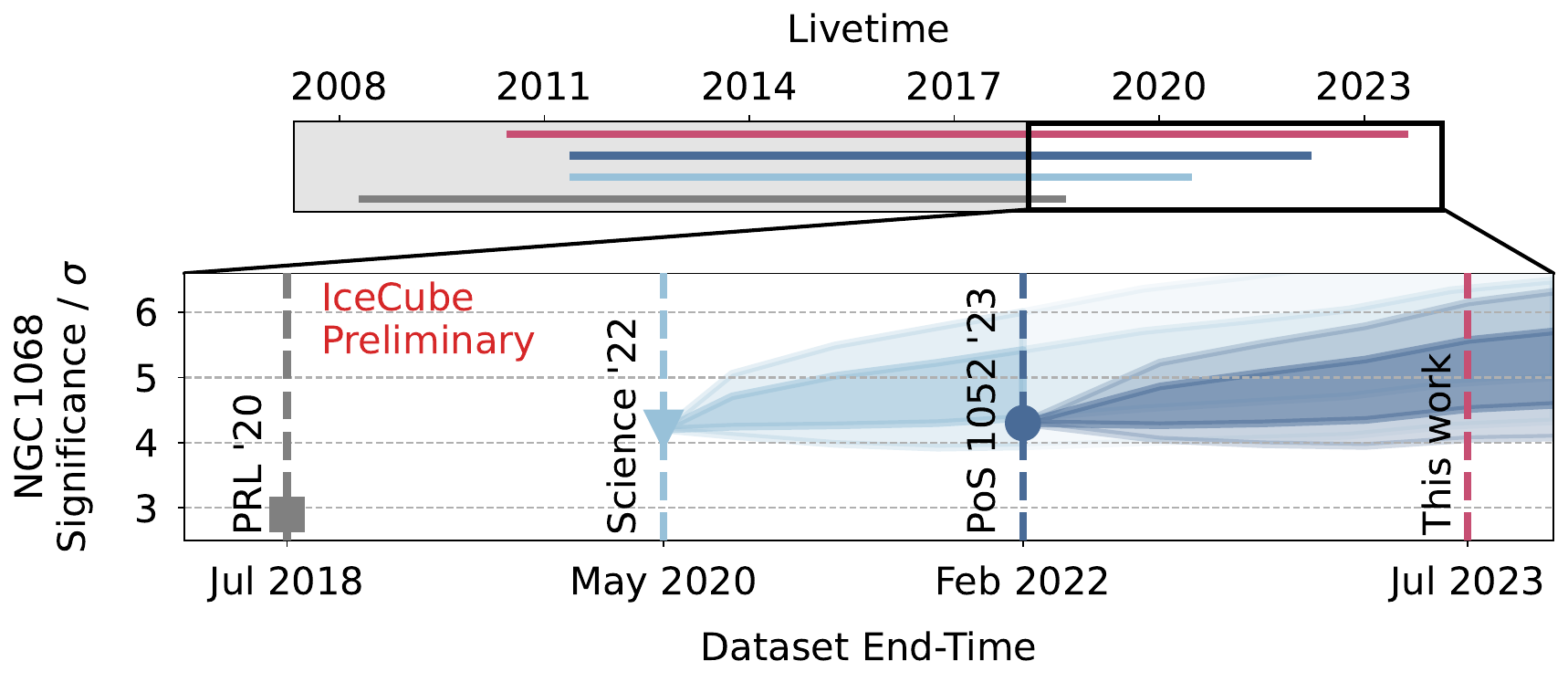}
    \caption{The upper panel compares the livetimes of the various datasets used to measure the neutrino flux from NGC~1068. The lower panel zooms on the end-time of the datasets and shows the increasing significance of the emission over the years. The result from Ref.~\cite{IceCubePRL2020} (grey square) used different data processing and analysis methods. Starting from Ref.~\cite{IceCubeScience2022} (light blue triangle), the rise in significance is due to the increased statistics. The significance from Ref.~\cite{PoS-ICRC2023-1052} (dark blue dot) is compatible with the expected statistical fluctuations within 1$\,\sigma$ (dark shaded bands). The lighter shaded bands indicate the 95\% containment of the fluctuations.}
    \label{fig:evolution}
\end{figure}

\noindent For NGC~1068, we evaluate the expected increase in significance.
It was detected for the first time at 2.9$\,\sigma$ when searching for astrophysical emission from 110 selected objects \cite{IceCubePRL2020}. This result used 10 years of data (2008 -- 2018) processed according to different standards depending on the detector configuration and analyzed with the previous methods. Out of the same search on 110 objects, the flux from NGC~1068 was measured again using the new processing and analysis methods on 8.7 years of data with a global significance of 4.2$\,\sigma$ \cite{IceCubeScience2022}. Finally, a new result was recently obtained by an analysis searching for emission from Seyfert Galaxies in the Northern sky with 10.4 years of the same data processing \cite{PoS-ICRC2023-1052}. The livetimes of the various datasets are compared in the upper panel of \autoref{fig:evolution}.
We can take the local significance of the latest work and calculate a global significance assuming that the same search of 110 candidate sources had been done. Doing so, we find a slightly increased significance of 4.3$\,\sigma$.
This result falls within the 68\% band of the expected statistical fluctuations, as can be seen in \autoref{fig:evolution}. The fluctuations are estimated by generating pseudo-experiments with 8.7 years of experimental data and simulating additional years of data by adding background and signal MC events according to
%adding simulated background and signal events according to
the best-fit steady flux measured in \cite{IceCubeScience2022}. We repeat the procedure starting from the result presented in \cite{PoS-ICRC2023-1052} to estimate the expected significance of NGC~1068 for this work (see \autoref{fig:evolution}), finding a 68\% chance to detect NGC~1068 between 4.5$\,\sigma$ and 5.5$\,\sigma$.

\section{Outlook: Improving the energy reconstruction with Graph Neural Networks}
\label{sec:GNN}
% Rasmus and Martin
\noindent We further investigate other approaches to improve the analysis sensitivity. One promising option is to improve the energy reconstruction using Graph Neural Networks (GNNs). GNNs are a class of machine learning algorithms that learn on graph-structured data. 
Graphs provide a strikingly simple but flexible data representation, allowing them to naturally represent the irregular geometry of neutrino detectors. 
As a preliminary study, we probe the agreement of predictions from DynEdge \cite{dynedge_paper}, the GNN from the open-source GNN library for neutrino telescopes \textit{GraphNeT} \cite{graphnet_paper}, in simulated and experimental data for IC79. We describe IceCube neutrino events as point cloud graphs, where each node represents an optical module that measured Cherenkov radiation during the neutrino interaction. Averages of arrival times and charges of reconstructed photon pulses are selected as node features. Such extreme aggregations are known to worsen the energy resolution but increase the agreement between simulation and observation. This choice can thus provide an upper bound on the agreement between simulated and experimental data and a lower bound on expected resolution. Nodes are connected to their 8 nearest neighbors as given by the Euclidean distance of the DOM positions.
\begin{wrapfigure}{r}{0.5\textwidth}
    \centering
    \includegraphics[width=0.50\textwidth]{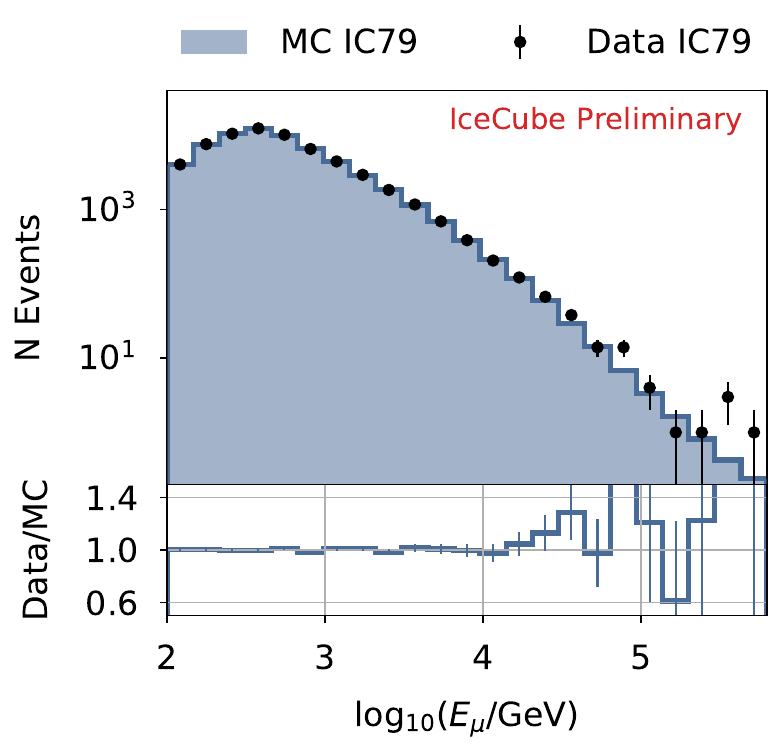}
    \caption{Agreement in the GNN energy estimation for the IC79 experimental data with the IC79 MC. The ratio between the experimental data and the MC is shown in the lower panel.}
    \label{fig:gnn_energy}
\end{wrapfigure}
We train DynEdge
%to reconstruct the muon energy
on a small sample of 1.2 million simulated track-like events. Choices in the loss function and training procedure are identical to \cite{dynedge_paper}.
As seen in Figure \ref{fig:gnn_energy}, the GNN can predict the muon energy with good agreement on both simulated and experimental data. The agreement is best in energy bins with a large number of training examples, and as the statistics decrease, so does the agreement. This trend is seen for the distribution of energies reconstructed with the DNN in Figure \ref{fig:ic79} as well. As for the resolution, we find that the GNN and the DNN achieve comparable performances at 1 TeV and below despite the GNN being trained on 1/6th of the data. However, above 1 TeV, resolutions are slightly worse, likely due to low statistics in the training sample. In the future, additional tests will be carried out with less aggressive aggregation of pulse information and a higher number of training examples. 

\section{Summary}
\noindent We have presented a future extension of the search for neutrino point sources in the Northern sky that found evidence for neutrino emission from the Seyfert II Galaxy NGC~1068. This work will include four additional years of observations. The statistical analysis relies on the excellent agreement between the experimental and the simulated data, which is now obtained for both, the complete IceCube detector configuration (IC86) and the incomplete one (IC79). %We obtained the desired level of agreement between experimental data and MC thanks to a reprocessing of the simulations that correctly mimics the IC79 detector response.
Because of this result, we can add 4 more years of observations to the analysis, including 3 years of data collected with the full IceCube detector and 1 year from the time before IceCube was completed. The increased statistics improve the sensitivity and discovery potential of the analysis by $30\%$ ($20\%$) for hard (soft) spectral emission. At the same time, the projected significance of the neutrino emission from NGC~1068, assuming it is a neutrino source, is expected to fall between 4.5$\,\sigma$ and 5.5$\,\sigma$ at a 68\% confidence level. Furthermore, we doubled the size of the simulations we use to construct the spatial and energy pdfs entering the likelihood formalism. We use the larger MC dataset to model the spatial clustering and energy distribution of events with energies larger than 100~TeV more accurately. We expect this improvement to reduce further the bias and the variance in the fit parameters and to improve the discovery potential for sources emitting hard-spectrum energy fluxes, as, for example, the blazar TXS~0506+056. Finally, we started investigating the improvement of the muon energy reconstruction by replacing our DNN with a GNN. Even with a relatively small training sample, we could obtain agreement between experimental data and simulations with a prototype GNN.
Further studies will clarify whether the GNN has the potential to provide an additional boost in sensitivity.

% Bibtex references:
% \setlength{\columnsep}{30pt}
% \begin{multicols}{2}
%     \bibliographystyle{ICRC}
%     \footnotesize
%     \bibliography{references}
% \end{multicols}

\bibliographystyle{ICRC}
% \footnotesize
\bibliography{references}

\providecommand{\href}[2]{#2}\begingroup\raggedright\begin{thebibliography}{10}

\bibitem{IceCubeScience2013}
{\bfseries IceCube} Collaboration, M.~G. Aartsen {\em et~al.}
  \href{http://dx.doi.org/10.1126/science.1242856}{{\em Science} {\bfseries
  342} no.~6161, (2013) 1242856}.

\bibitem{TXSScience2018}
{\bfseries IceCube, Fermi-LAT, MAGIC, AGILE, ASAS-SN, HAWC, H.E.S.S., INTEGRAL,
  Kanata, Kiso, Kapteyn, Liverpool Telescope, Subaru, Swift NuSTAR, VERITAS,
  and VLA/17B-403} Collaboration, M.~Aartsen {\em et~al.}
  \href{http://dx.doi.org/10.1126/science.aat1378}{{\em Science} {\bfseries
  361} no.~6398, (2018) eaat1378}.

\bibitem{IceCubeScience2018}
{\bfseries IceCube} Collaboration, M.~G. Aartsen {\em et~al.}
  \href{http://dx.doi.org/10.1126/science.aat2890}{{\em Science} {\bfseries
  361} no.~6398, (2018) 147--151}.

\bibitem{IceCubeApJ2009}
{\bfseries IceCube} Collaboration, R.~Abbasi {\em et~al.}
  \href{http://dx.doi.org/10.1088/0004-637X/701/1/L47}{{\em The Astrophysical
  Journal Letters} {\bfseries 701} no.~1, (2009) L47--L51}.

\bibitem{IceCubePRL2020}
{\bfseries IceCube} Collaboration, M.~G. Aartsen {\em et~al.}
  \href{http://dx.doi.org/10.1103/PhysRevLett.124.051103}{{\em Phys. Rev.
  Lett.} {\bfseries 124} no.~5, (2020) 051103}.

\bibitem{BRAUN2008299}
J.~Braun {\em et~al.}
  \href{http://dx.doi.org/https://doi.org/10.1016/j.astropartphys.2008.02.007}{{\em
  Astroparticle Physics} {\bfseries 29} no.~4, (2008) 299--305}.

\bibitem{IceCubeScience2022}
{\bfseries IceCube} Collaboration, R.~Abbasi {\em et~al.}
  \href{http://dx.doi.org/10.1126/science.abg3395}{{\em Science} {\bfseries
  378} no.~6619, (2022) 538--543}.

\bibitem{IceCubeJInst2017}
{\bfseries IceCube} Collaboration, M.~G. Aartsen {\em et~al.}
  \href{http://dx.doi.org/10.1088/1748-0221/12/03/P03012}{{\em Journal of
  Instrumentation} {\bfseries 12} no.~03, (2017) P03012--P03012}.

\bibitem{IceCubeDiffuseAPJ2022}
{\bfseries IceCube} Collaboration, R.~Abbasi {\em et~al.}
  \href{http://dx.doi.org/10.3847/1538-4357/ac4d29}{{\em The Astrophysical
  Journal} {\bfseries 928} no.~1, (2022) 50}.

\bibitem{Poluektov:2014rxa}
A.~Poluektov \href{http://dx.doi.org/10.1088/1748-0221/10/02/P02011}{{\em
  Journal of Instrumentation,} {\bfseries 10} no.~02, (2015) P02011}.

\bibitem{PoS-ICRC2023-1052}
{\bfseries IceCube} Collaboration, T.~Glauch, A.~Kheirandish, T.~Kontrimas,
  Q.~Liu, and H.~Niederhausen {\em PoS} {\bfseries ICRC2023} (these
  proceedings) 1052.

\bibitem{dynedge_paper}
{\bfseries IceCube} Collaboration, R.~Abbasi {\em et~al.}
  \href{http://dx.doi.org/10.1088/1748-0221/17/11/P11003}{{\em JINST}
  {\bfseries 17} no.~11, (2022) P11003}.

\bibitem{graphnet_paper}
A.~Søgaard, R.~F. Ørsøe, M.~Holm, L.~Bozianu, A.~Rosted, T.~C. Petersen,
  K.~E. Iversen, A.~Hermansen, T.~Guggenmos, P.~Andresen, M.~H. Minh, L.~Neste,
  M.~Holmes, A.~Pontén, K.~L. DeHolton, and P.~Eller
  \href{http://dx.doi.org/10.21105/joss.04971}{{\em Journal of Open Source
  Software} {\bfseries 8} no.~85, (2023) 4971}.

\end{thebibliography}\endgroup

% Alternatively, you can include references by hand:
%\begin{thebibliography}{99}
%\bibitem{...}
%
%\end{thebibliography}

\clearpage

%The following list of authors, affiliations and funding agencies will be updated at the day of submission. The following template is a placeholder generated via https://authorlist.icecube.wisc.edu/icecube on March 25, 2023 and will be updated.
\section*{Full Author List: IceCube Collaboration}

\scriptsize
\noindent
R. Abbasi$^{17}$,
M. Ackermann$^{63}$,
J. Adams$^{18}$,
S. K. Agarwalla$^{40,\: 64}$,
J. A. Aguilar$^{12}$,
M. Ahlers$^{22}$,
J.M. Alameddine$^{23}$,
N. M. Amin$^{44}$,
K. Andeen$^{42}$,
G. Anton$^{26}$
% ...
\\
\\
% ...
$^{12}$ Universit{\'e} Libre de Bruxelles, Science Faculty CP230, B-1050 Brussels, Belgium \\
% ...
$^{17}$ Department of Physics, Loyola University Chicago, Chicago, IL 60660, USA \\
$^{18}$ Dept. of Physics and Astronomy, University of Canterbury, Private Bag 4800, Christchurch, New Zealand \\
% ...
$^{22}$ Niels Bohr Institute, University of Copenhagen, DK-2100 Copenhagen, Denmark \\
$^{23}$ Dept. of Physics, TU Dortmund University, D-44221 Dortmund, Germany \\
% ...
$^{26}$ Erlangen Centre for Astroparticle Physics, Friedrich-Alexander-Universit{\"a}t Erlangen-N{\"u}rnberg, D-91058 Erlangen, Germany \\
% ...
$^{40}$ Dept. of Physics and Wisconsin IceCube Particle Astrophysics Center, University of Wisconsin{\textendash}Madison, Madison, WI 53706, USA \\
% ...
$^{42}$ Department of Physics, Marquette University, Milwaukee, WI, 53201, USA \\
% ...
$^{44}$ Bartol Research Institute and Dept. of Physics and Astronomy, University of Delaware, Newark, DE 19716, USA \\
% ...
$^{63}$ Deutsches Elektronen-Synchrotron DESY, Platanenallee 6, 15738 Zeuthen, Germany  \\
$^{64}$ Institute of Physics, Sachivalaya Marg, Sainik School Post, Bhubaneswar 751005, India \\
% ...

\subsection*{Acknowledgements}

\noindent
The authors gratefully acknowledge the support from the following agencies and institutions:
USA {\textendash} U.S. National Science Foundation-Office of Polar Programs,
U.S. National Science Foundation-Physics Division,
U.S. National Science Foundation-EPSCoR,
Wisconsin Alumni Research Foundation,
Center for High Throughput Computing (CHTC) at the University of Wisconsin{\textendash}Madison,
Open Science Grid (OSG),
Advanced Cyberinfrastructure Coordination Ecosystem: Services {\&} Support (ACCESS),
Frontera computing project at the Texas Advanced Computing Center,
U.S. Department of Energy-National Energy Research Scientific Computing Center,
Particle astrophysics research computing center at the University of Maryland,
Institute for Cyber-Enabled Research at Michigan State University,
and Astroparticle physics computational facility at Marquette University;
Belgium {\textendash} Funds for Scientific Research (FRS-FNRS and FWO),
FWO Odysseus and Big Science programmes,
and Belgian Federal Science Policy Office (Belspo);
Germany {\textendash} Bundesministerium f{\"u}r Bildung und Forschung (BMBF),
Deutsche Forschungsgemeinschaft (DFG),
Helmholtz Alliance for Astroparticle Physics (HAP),
Initiative and Networking Fund of the Helmholtz Association,
Deutsches Elektronen Synchrotron (DESY),
and High Performance Computing cluster of the RWTH Aachen;
Sweden {\textendash} Swedish Research Council,
Swedish Polar Research Secretariat,
Swedish National Infrastructure for Computing (SNIC),
and Knut and Alice Wallenberg Foundation;
European Union {\textendash} EGI Advanced Computing for research;
Australia {\textendash} Australian Research Council;
Canada {\textendash} Natural Sciences and Engineering Research Council of Canada,
Calcul Qu{\'e}bec, Compute Ontario, Canada Foundation for Innovation, WestGrid, and Compute Canada;
Denmark {\textendash} Villum Fonden, Carlsberg Foundation, and European Commission;
New Zealand {\textendash} Marsden Fund;
Japan {\textendash} Japan Society for Promotion of Science (JSPS)
and Institute for Global Prominent Research (IGPR) of Chiba University;
Korea {\textendash} National Research Foundation of Korea (NRF);
Switzerland {\textendash} Swiss National Science Foundation (SNSF);
United Kingdom {\textendash} Department of Physics, University of Oxford.

\end{document}